# Dry electrodes for bioimpedance measurements – design, characterization and comparison

R Kusche[1], S Kaufmann[1], and M Ryschka[1]
[1] Laboratory of Medical Electronics (LME), Lübeck University of Applied Sciences, Lübeck, Germany

E-mail: roman.kusche@fh-luebeck.de, and martin.ryschka@fh-luebeck.de

**Abstract.** *Objective:* Bioimpedance measurements are mostly performed utilizing gel electrodes to decrease the occurring electrode-skin impedance. Since in many measurement environments this kind of electrode is not appropriate, the usability of dry electrodes is analysed. *Approach:* The development of five different kinds of dry electrodes, including gold, stainless steel, carbon rubber and metallized textile as contact materials are proposed. All test electrodes are based on a circular printed circuit board as carrier and have the same contact surface dimensions. To compare the electrodes' characteristics, the occurring electrode-skin impedances are measured under variation of signal frequency, contact duration, contact pressure, placement position and subjects. Additionally, all measurements are performed with silver/silver chloride (Ag/AgCl) dry gel electrodes for comparison purposes. *Main results:* The analysed parameters play a significant role regarding the electrode-skin impedance. Choosing a wise setup of these parameters can decrease the electrode-skin impedance of dry electrodes down to ranges of dry gel electrodes and even below. *Significance:* The usage of dry electrodes is one of the most difficult challenges when transferring scientific measurement techniques to clinical environments or commercial products but it is indispensable for many applications like body composition measurements or prosthesis control.

## 1. Introduction

Bioimpedance measurements are a well-known technique to determine the electrical characteristic of living tissue (Grimnes and Martinsen 2008). It is commonly used to analyse the human body composition (Matthie 2008) or for monitoring time-dependent biosignals. These signals can for example contain information about respiration (Martinsen *et al* 2014, Kusche *et al* 2015), the cardiovascular system (Nyboer 1950, González Landaeta *et al* 2006, Kusche *et al* 2018b) or the contractions of muscles (Rutkove 2009).

To measure the bioimpedance, most published devices (Ferreira *et al* 2017, Honero *et al* 2013, Kaufmann *et al* 2014, Kusche *et al* 2018a) apply a known alternating current (AC) of some Milliamperes (mA) in a frequency range of tens or hundreds of Kilohertz (kHz) to the tissue of interest via two electrodes. Simultaneously, a second pair of electrodes acquires the occurring voltage drop over the bioimpedance. The ratio of measured voltage to applied current yields to the complex impedance value containing the magnitude and phase values. This method differs significantly from other bioelectri-cal measurement techniques like electrocardiography (ECG), electromyography (EMG) or electroencephalography (EEG), in which there is almost no current flowing between the subject and the electrical instrumentation. A major issue regarding the used excitation current is that for flowing through the bioimpedance, the current has to pass two electrode-skin interfaces as well. Depending on the specific measurement setup and the used electrodes, the magnitude of these electrode-skin impedances ($Z_{ES}$) can be much higher than the actual bioimpedance of interest (Rosell *et al* 1988, Grimnes 1983, Grimnes and Martinsen 2008). Even when the influence of this effect to the measurement results can be reduced by specific setups, like the four-terminal sensing, it is one of the most challenging problems in the field of bioimpedance instrumentation (Grimnes and Martinsen 2008). The occurring voltage drops over these electrode-skin impedances, when applying a current to the tissue, can for example exceed the compliance voltage of the current source circuit or the admissible voltage range of the measurement circuitry.

In many bioimpedance applications, non-polarizing silver/silver chloride (Ag/AgCl) gel electrodes are utilized to keep this issue under control. Unfortunately, these electrodes are commonly single-use parts, which are stuck to the skin by an adhesive electrode surface. This adhesive complicates improving the electrodes positioning, once the electrodes are attached and can also cause skin irritations (Baek *et al* 2008). Additionally, the removal of the electrodes after a measurement can be painful to the subject. Neither suction electrodes can solve the problem, since skin irritation due to the necessary negative pressure increases with application time and may cause severe skin damage. In many real measurement environments, these kinds of electrodes are not appropriate. These environments can for instance be a limb prosthesis, controlled via bioimpedance signals (Rutkove 2009), or commercial body-fat scales (Jebb *et al* 2000).

An alternative is the usage of reusable dry electrodes, which are not adhesive and therefore have to be pressed to the skin via an external force. These electrodes are commonly made of metal or carbonized rubber but do not provide an electrolyte (Searle and Kirkup 2000, Pylatiuk *et al* 2009, Meziane *et al* 2013, Chi *et al* 2010). In the past several years, textile electrodes have become popular as well, probably caused by the increasing interest in wearable electronics (Colyer and McGuigan 2018, Ankhili *et al* 2018, Pani *et al* 2015).

The design, as well as the characterization of several dry electrodes have been published in the past (Chi *et al* 2010, Laferriere *et al* 2011, Chan and Lemaire 2010). However, these publications typically consider ECG, EMG or EEG applications and do therefore not focus on the electrode-skin impedances.

In this work, the design of five different types of dry electrodes is proposed, realizing gold, stainless steel, smooth carbonized rubber, textured carbonized rubber and conductive textile as contact materials. After measuring the impedances of only the electrodes, the actual electrode-skin impedances are determined under several conditions in a frequency range from 24 kHz to 391 kHz. The considered parameters are time-, force- and subject-dependency. For this measurements four healthy volunteers were recruited.

## 2. Materials and methods

### 2.1. Electrode-Skin Impedance

Since both, the electrode and the skin influence each other, the electrical behaviour of the electrode-skin interface cannot be described as a simple series connection of electrode impedance and skin impedance (Chi *et al* 2010). Therefore, only analysing the combination of both is useful.

In the past, several equivalent circuits to model the electrical behaviour of electrode-skin interfaces have been published (Gruetzmann *et al* 2007, Lin *et al* 2011, Chi *et al* 2010). Since this work does not focus on the occurring half-cell voltage or other effects, but exclusively on the electrode-skin impedance, only the passive electrical behaviour is modelled.

The resulting equivalent circuit for the electrode-skin interface of a dry electrode is based on other published models (Grimnes and Martinsen 2008, Chi *et al* 2010) and shown in figure 1. It consists of the combination of four conductive layers. The first layer represents the electrode itself, which has a resistive behaviour ($R_{El}$) depending on the specific setup and used materials. The second layer is the actual interface between the electrode and the skin. A direct contact between both, is modelled as a contact resistance $R_C$. Since the skin and the electrode do not yield to an exact form closure, air inclusions can occur, which lead to an additional capacitive behaviour, represented by the contact capacity $C_C$. It is conceivable that this effect depends significantly on time, skin preparation and other factors. The third layer represents the epidermis and is also represented by a parallel connection of a resistance $R_{Ep}$ and a capacity $C_{Ep}$. Typically, the dermis is modelled to be the fourth layer with a resistive electrical behaviour $R_D$.

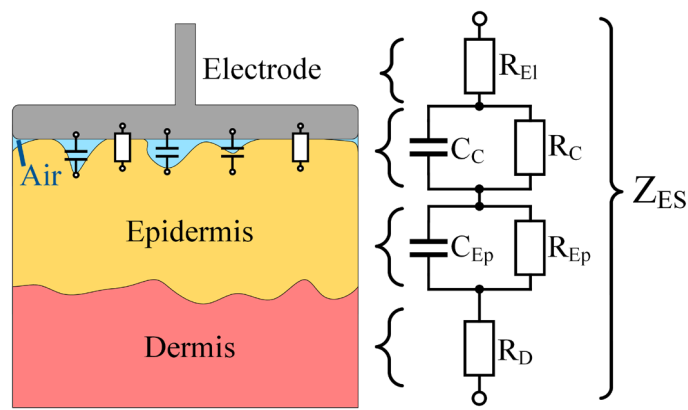

**Figure 1.** Electrical equivalent circuit of a dry electrode connected to the skin. The electrode-skin impedance $Z_{ES}$ includes the electrode resistance $R_{El}$, the contact impedance $C_C \| R_C$, the epidermis impedance $C_{Ep} \| R_{Ep}$ and the resistance of the dermis $R_D$.

Since for impedance measurements, at least two electrodes are necessary, in this work the total impedance magnitude $|Z_{total}|$, composed of two electrode-skin impedances ($Z_{ES1}$, $Z_{ES2}$) and the bioimpedance $Z_{Bio}$ in-between, is measured:

$$|Z_{total}| = |Z_{ES1} + Z_{Bio} + Z_{ES2}| \qquad (1)$$

To apply the excitation current, it has to pass two electrode-skin transitions, forming the significant burden of the current source. So the total impedance is the more relevant value in comparison to a single electrode's impedance. Former publications have shown that the magnitude values of $Z_{Bio}$ are much lower than the electrode-skin impedances $|Z_{ES1}|$ and $|Z_{ES2}|$ (Dastjerdi *et al* 2013) and thus may be neglected.

*2.2. Developed Dry Electrodes*

To insure comparability to the largest extent, for this work, five different kinds of dry electrodes have been developed. All of them are based on a circular printed circuit board (PCB) with a diameter of $D = 15$ mm and a thickness of $t = 1.55$ mm. Hence, the resulting electrode surface is $A = 177$ mm² for all electrodes. To enable the connection to a measurement cable, a gold-plated pin is soldered to the top side of the PCB. The specific build-up of each electrode, shown in figure 2, is described briefly:

(1) Gold-plated:
   The bottom side of the PCB consists of gold-plated copper.
(2) Stainless Steel:
   This build-up is based on the gold electrode. Additionally, a layer of stainless steel with a thickness of 0.3 mm is attached to the gold surface, using a conductive adhesive. This silver epoxy adhesive (8330S, MG Chemicals, Surrey, CA) has a resistivity of 7 Ω·mm²/m and is used for the following electrodes, as well. The stainless steel surface has been pre-scrubbed utilizing a sandpaper with a P280 IOS/FEPA grit size.

(3) Carbon Rubber (smooth surface):
The contact surface consists of carbonized rubber (Pierenkemper GmbH, Ehringshausen, DE) and is attached via the previously described conductive adhesive. The surface is pre-scrubbed in the same manner as the stainless steel electrode.

(4) Carbon Rubber (textured surface):
Textured carbon rubber (VITAtronic Ltd., Dornstetten, DE) is used as contact material. It is attached via the conductive epoxy adhesive, as well.

(5) Conductive Textile:
There are several conductive textiles available in the market, but many are intended for shielding purposes and do not provide very high conductivities. The used commercially available woven conductive fabric (Art.No. 1168, Adafruit, New York City, NY, US) consists of a conductive layer with a surface resistivity of ≤50 mΩ/square. and an isolating layer. Therefore, the textile is wrapped around an adhesive tape before attaching it to the carrier PCB, as shown in figure 2.

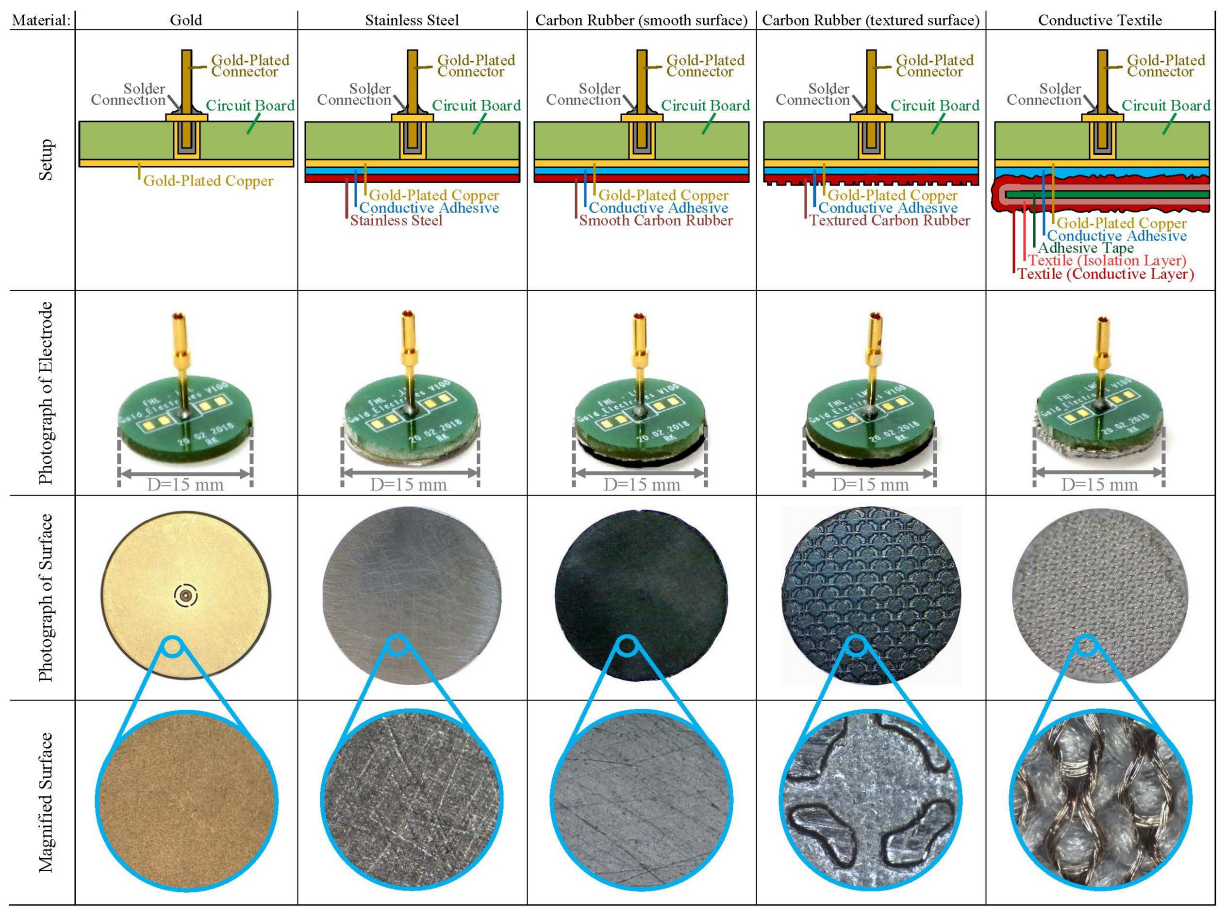

**Figure 2.** Setup and photographs of the developed dry electrodes. All electrodes consist of a circular PCB as carrier and an additional contact material, which is attached via a conductive epoxy adhesive. Photographs of the contact surfaces are depicted as well. In the bottom row, the surfaces are magnified by a factor of 7.5 by a microscope.

### 2.3. Electrodes Sleeve

For comfortable positioning of the non-adhesive dry electrodes and to ensure a defined distance between the electrodes under test, an elastic sleeve (Vertics.Sleeves Size 3, Vertics, Wiesbaden, DE) is utilized. It also ensures constant and comfortable contact forces of approximately 1 N.

In figure 3, two photographs are shown. The upper one illustrates the sleeve, which is equipped with two electrodes of each kind. In this case, all electrodes are positioned on the upper-side of the forearm. Below, a second setup is shown to position the electrodes on the forearm's under-side.

In addition to the previously described dry electrodes, an additional pair of Ag/AgCl dry gel electrodes (Kendall H92SG, Medtronic, Minneapolis, MN, USA) for comparison purposes is equipped. Its surface diameter is customized to be the same as the diameter of the dry electrodes.

The electrodes of each pair have distances of d = 2.5 cm. For better visualisation, the electrodes in figure 3 are numbered ($E_1$: Ag/AgCl; $E_2$: Gold; $E_3$: Carbon Rubber (smooth surface); $E_4$: Stainless Steel; $E_5$: Carbon Rubber (textured surface); $E_6$: Metallized Textile).

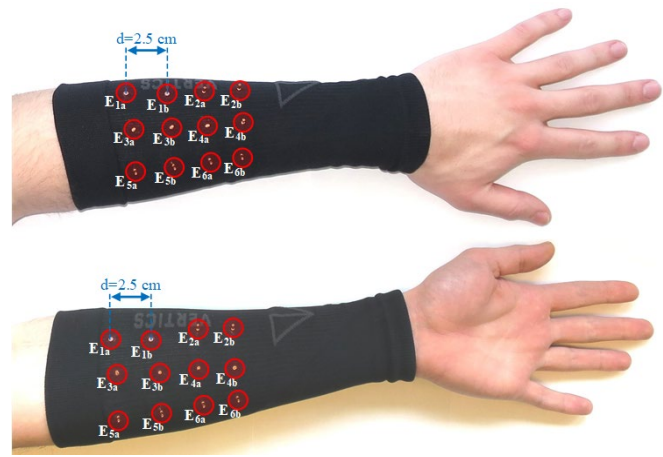

**Figure 3.** Sleeve for equidistant electrode pair positioning. In the upper photograph, the setup for positioning the electrode pairs to the upper-side of the forearm is shown. Below, the setup to place the electrodes at the under-side is depicted. Five pairs of dry electrodes and one pair of Ag/AgCl dry gel electrodes are equipped ($E_1$: Ag/AgCl; $E_2$: Gold; $E_3$: Carbon Rubber (smooth surface); $E_4$: Stainless Steel; $E_5$: Carbon Rubber (textured surface); $E_6$: Metallized Textile). The red circles mark the contact surfaces of the electrodes below the sleeve.

### 2.4. Impedance Measurements

The measurement of the impedances are performed by a previously published high accuracy bioimpedance measurement system (Kaufmann *et al* 2014). It is capable of acquiring 3840 complex impedance spectra per second in a frequency range from 24 kHz to 391 kHz. The excitation current can be chosen to be between 10 µA and 5 mA and is set in this work to 10 µA. This configuration leads to a system specific impedance measurement range of 37 kΩ. To connect the device to the electrodes under test, twisted pair cables with a length of 40 cm are utilized. Due to the electrical safety requirements, the device is powered by an external medical power supply.

## 3. Results

### 3.1. Frequency Responses of Electrodes

Before analysing the total impedances, the behaviour of only the electrodes is determined. For this purpose the setup, shown in figure 4, is used. Two electrodes of the same kind are aligned to each other. To generate realistic and reproducible measurement conditions, a force of F =1 N is applied to this setup and verified by a force gauge (FG-5000A-232, Lutron Electronic, Taipei, Taiwan).

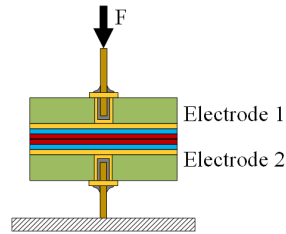

**Figure 4.** Measurement setup to determine the electrode impedances. Electrode 1 and Electrode 2 are equal and their conductive surfaces are aligned. A force of F = 1 N is applied.

The impedance magnitude $|Z_E|$ of this series connection of two electrodes is measured over a frequency range from 24 kHz to 391 kHz by the previously described bioimpedance measurement device.

In figure 5, the measurement results are shown. It can be seen that the electrode impedances are very stable over frequency. Just the plot of the gold electrode changes visibly over frequency. Since this change is in m$\Omega$ ranges and has an inductive behaviour, it is assumed to be caused by the measurement cables. Nevertheless, it is the electrode with the lowest impedance over the whole frequency range.

The stainless (SL) steel electrode and the textile electrode have similar impedance values of about 1 $\Omega$. It is plausible that their impedance is slightly higher than that of the gold electrode, since both contain an additional layer of conductive adhesive and the corresponding contact interfaces.

The impedances of the rubber electrodes are in a range of tens of Ohms and the pair of Ag/AgCl dry gel electrodes leads to the highest impedance of about 100 $\Omega$.

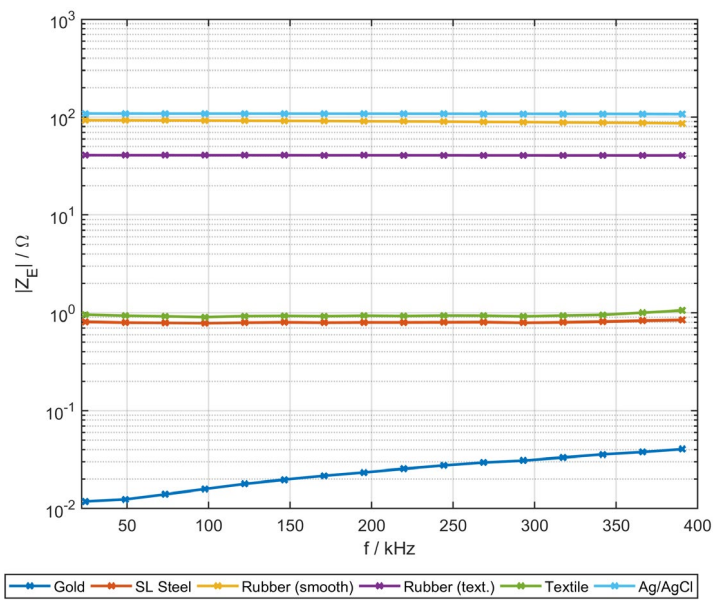

**Figure 5.** Measured frequency responses of the five dry electrodes and the Ag/AgCl dry gel electrodes in a frequency range from 24 kHz to 391 kHz. The impedances $|Z_E|$ are measured as shown in figure 4 and do therefore represent the series connection of two equal electrodes.

Even when this result demonstrates that dry electrodes have lower impedances than the Ag/AgCl dry gel electrode, it does not represent the actual electrode-skin impedance. This is typically in k$\Omega$ ranges and thus much more significant than just the electrodes' impedances.

*3.2. Time Dependency of the Electrode-Skin Impedances*

Since dry electrodes do not provide an electrolyte, primarily the skin conditions affect the electrode-skin impedance. Therefore, after applying the electrodes to the dry skin, the capacitive behaviour is very dominant. After a certain time, the accumulation of fluid under the electrodes works as electrolyte and the electrode-skin interface becomes less capacitive and more resistive (Kaufmann *et al* 2013).

This effect is analysed by performing simultaneous measurements of all six kinds of electrodes utilizing the previously described sleeve. One hour before the actual measurement, the skin of the young male subject was washed with soap and the electrodes were cleaned with ethanol. The electrodes are positioned on the upper-side of the subject's forearm and the total impedance is measured over time and frequency for a duration of 30 minutes (T = {0, 1, 2, 3, 4, 5, 10, 15, 20, 25, 30} min). The contact forces of the electrodes have been measured by a force gauge (FG-5000A-232, Lutron Electronic, Taipei, Taiwan) to be approximately 1 N.

In figure 7, the results of this measurement are shown. The total impedances of all dry electrode pairs, except the textile electrode, decrease over time. As expected, the Ag/AgCl dry gel electrodes generate a low impedance of about 1 kΩ, almost constant over the examined time frame.

Both metal electrodes provide a very similar behaviour. Within the first 5 minutes their total impedances decrease by half and after about 15 minutes, the impedances go to a saturation state. The carbonized rubber electrodes generate significantly higher impedances. Compared to the metal electrodes, the measured values of the smooth surface rubber electrode are about three times higher for the whole duration and all frequencies. The textured surface electrodes yield to impedances, which are even five times higher than those of metal electrodes.

The textile electrode is very independent regarding time. In the figure only the related plot for f = 391 kHz is depicted, in which the total impedance is about 15 kΩ. The usage of lower frequencies lead to electrode-skin impedances above the measurement range.

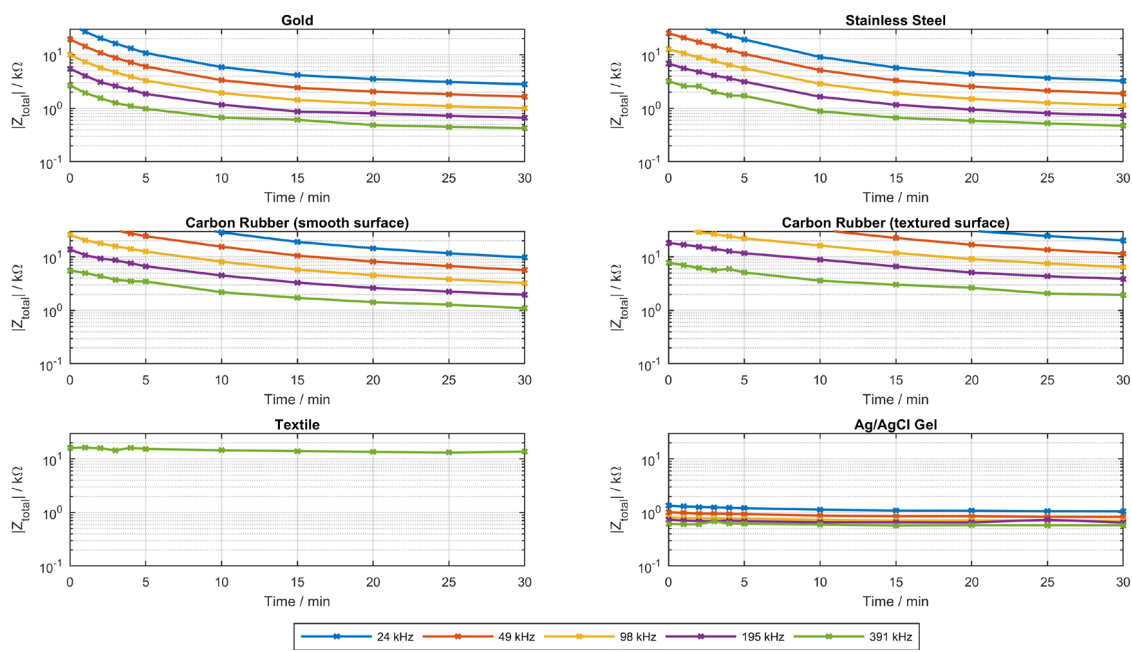

**Figure 6.** Time dependencies of the investigated total impedances $|Z_{total}|$, which contain two electrode-skin interface impedances and the bioimpedance in-between them. The measurement was performed on the upper-side of the subject's forearm. For better visualisation, the impedance magnitude is shown in the range from 100 Ω to 30 kΩ. The frequency ranges from 24 kHz to 391 kHz. In the plot of the textile electrode, just the results for f=391 kHz are shown, since the remaining results are not visible within these axes ranges.

Carbon rubber electrodes show higher impedance values than metal electrodes. An additional reduction of the actual electrode surface which is in contact with the skin, leads to the even higher impedances of the textured carbon rubber electrodes. The textile electrode's independency regarding time can be explained by the fact, that the occurring fluids are lead away from the skin by the textile, itself. It's very high impedance in general, is caused by the small effective conductive surface.

*3.3. Force Dependency of the Electrode-Skin Impedances*

The uneven surface of the skin and the force of the hairs against the electrodes can influence the resulting electrode-skin impedances. Therefore, one may expect that a higher force of the dry electrodes to the skin leads to lower impedance values. Since high forces can lead to discomfort of the subject, useful values of course are limited.

In figure 7, the principal measurement setup to determine the force dependency of the electrodes is shown. Two electrodes of the same kind are mounted on a common rigid carrier with a distance of d = 2.5 cm. The force, applied to this carrier, is split into two almost same forces F, which affect to the skin.

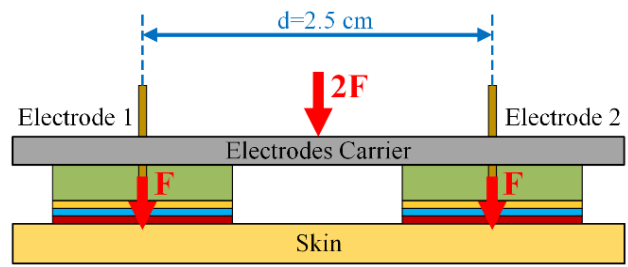

**Figure 7.** Principle measurement setup to analyse the influence of the contact force. Both electrodes are mounted to a common rigid electrodes carrier with a distance of d = 2.5 cm.

One hour before applying the electrodes to the subject's upper-side of the forearm, the skin of the young male subject was washed with soap and the electrodes were cleaned with ethanol. For more realistic results, the electrodes are attached 5 minutes before performing the actual measurement. For this measurement, the force F is varied in a range from 1 N to 20 N and the resulting total impedances are determined by the referenced bioimpedance measurement system.

In figure 8, the obtained impedance values for each pair of electrodes are plotted. It can be seen that the electrical contact between the electrode and the skin becomes better, when increasing the force to the dry electrodes. The major gain in conductivity is achieved by applying forces of up to 5 N. A further increase of the force does not change the electrode-skin impedance significantly.

In contrast to the measurements, shown before in figure 6, the textured rubber electrodes provide slightly lower impedance values than the rubber electrodes with a smooth surface. The reason for that could be, that when analysing the force dependencies, electrodes of different kinds have not been measured simultaneously. Therefore, the measurement conditions, especially the skin conditions cannot simply be compared with each other. It is also conceivable that the bigger surface of the textured electrodes, caused by the additional vertical surfaces, comes into contact with the skin, when a certain force is applied.

As shown before in the simplified equivalent circuit in Fig. 1, the electrode skin impedance depends on many physical factors, which are not analysed more detailed in this work.

In this experiment, also the impedance values of the textile electrodes reach the upper part of the measurement range, when applying high forces.

The total impedance of the Ag/AgCl dry gel electrodes is almost independent regarding the applied forces and decreases just slightly.

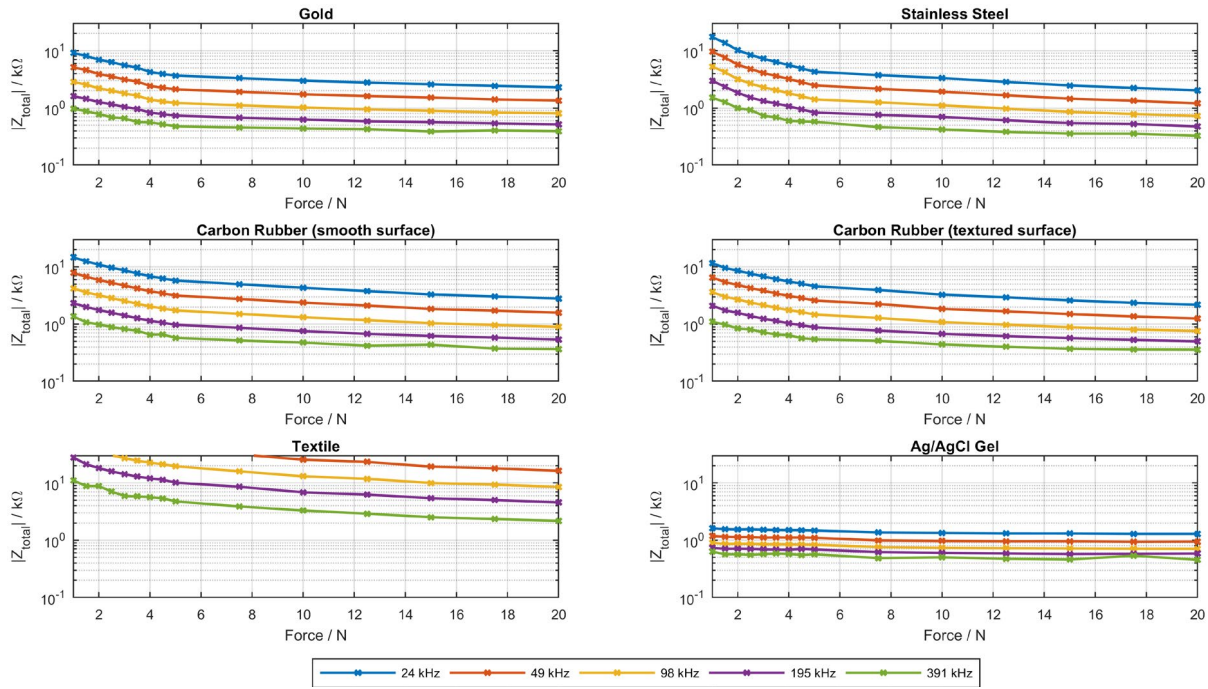

**Figure 8.** Measured total impedances $|Z_{total}|$ over contact force in a range from 1 N to 20 N under variation of the signal frequency from 24 kHz to 391 kHz.

This measurement shows that forces of up to 5 N can help to decrease the electrode-skin impedance significantly. Higher forces can reduce it even more, but to prevent discomforts, these forces are not applicable in most applications.

Especially, the combination of relatively high forces and high frequencies could be interesting for many applications. For instance, the usage of a frequency of 391 kHz and a force of 5 N leads for both the metal and the rubber electrodes to impedance values in the same range as the Ag/AgCl dry gel electrodes.

### 3.4. Subject Dependency

A major problem of electrode studies is that the electrode-skin impedances have a significant variation regarding electrode positions and measurement subjects. Minor differences in skin moisture or hairiness but also the thickness of the skin can lead to completely different conditions. Additionally, perspiration is a very complex and not predictable factor (Grimnes and Martinsen 2008, Yamamoto and Yamamoto 1977).

Therefore, the approach of this section is to determine the electrical characteristics of the electrodes under variation of the subjects and the electrode positions. In this study, measurements are performed on four (f=1, m=3) young subjects at two different electrode positions, the upper-side and the under-side of the forearm.

One hour before each measurement, the skin was washed with soap and the electrodes were cleaned with ethanol. The electrodes are placed 10 minutes before the actual measurement procedure with the help of the proposed sleeve, shown in figure 3. The resulting electrode contact forces are about 1 N.

In figure 9, the obtained frequency responses of the total impedances $|Z_{total}|$ are depicted.
It can be seen that even the small set of measurement data results into widely varying results. Only the impedance of the Ag/AgCl dry gel electrodes measurement leads to very constant values.

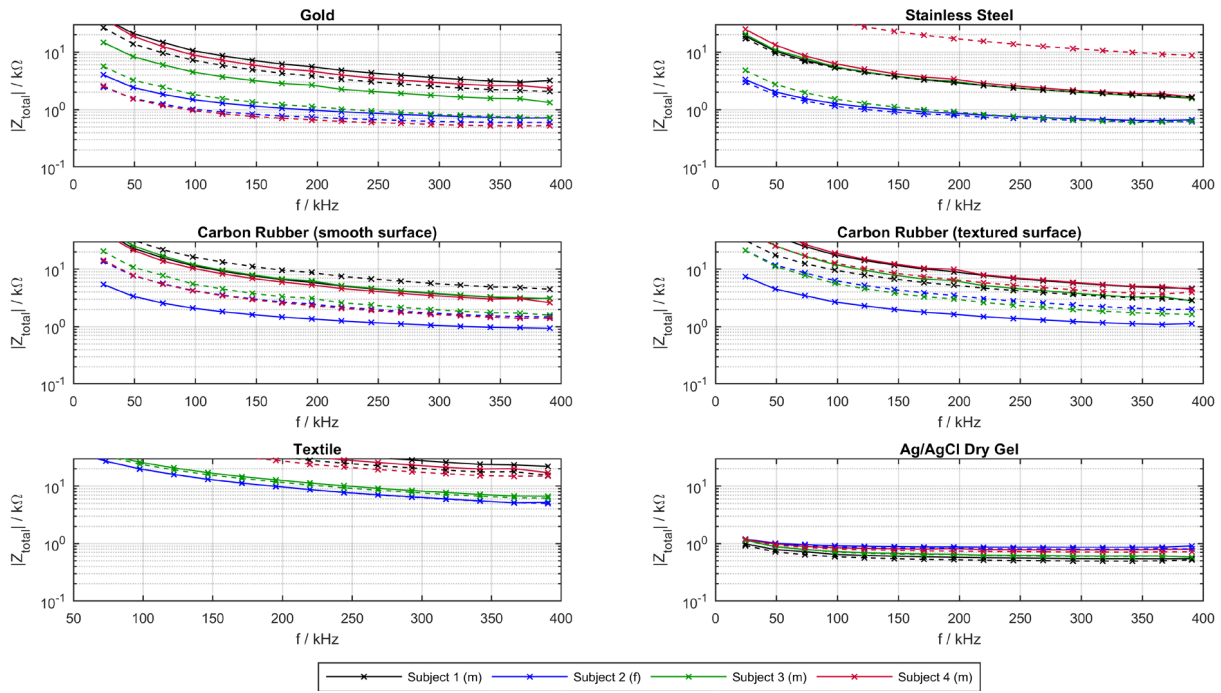

**Figure 9.** Frequency responses of the total impedances |$Z_{total}$| under variation of the subject and the electrodes position. The measurement was performed on 4 subjects each once on the forearm upper-side (solid lines) and once on the under-side (dashed lines).

It depends on every specific measurement application, if the average or the maximum impedance values are of interest. Especially in clinical environments the reliability of measurements has to be very high and therefore the maximum total impedance values might be of major interest. For comparison purposes, the mean values are more useful.

Thus, the averaged measurements are plotted into a common plot in figure 10. As already previously figured out, the Ag/AgCl dry gel electrodes lead to a very low and constant impedance over frequency. The dry electrode with the lowest total impedance values is the gold electrode, followed by the stainless steel electrode. Similar to the previous measurements, the carbon rubber electrodes' impedance values are higher than those of the metal electrodes. Especially the textured electrodes lead to high impedances. The impedance values of the textile electrodes are the highest by far.

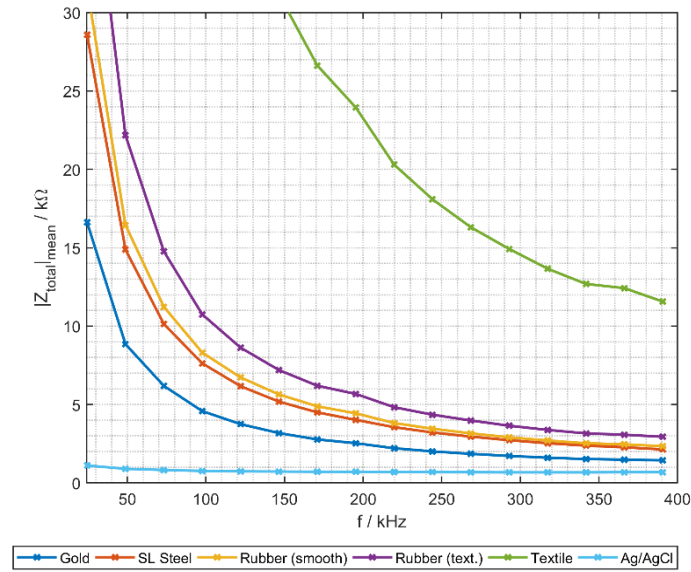

**Figure 10.** Averaged frequency responses of the eight measurements on the forearm, performed on the four subjects in a frequency range from 24 kHz to 391 kHz. As expected the Ag/AgCl electrodes offer the best performance over subjects and frequency. The best dry electrode performance is offered by the gold-plated electrodes.

### 4. Discussion

The measurements have shown that the occurring total impedances for the observed frequencies are in kΩ ranges. These high values make the influence of the electrodes' impedances, which have been measured to be tens of Ohms, negligible. For instance, the Ag/AgCl dry gel electrodes have the highest impedances of all proposed electrodes. However, when contacted to the skin, this electrode shows the lowest total interface impedances under almost all conditions.

When observing the occurring total impedances over frequency, like in figures 6-10, a predominant capacitive behaviour can be recognized, which corresponds to the equivalent circuit in figure 1.

The electrode-skin impedances of all compared dry electrodes depend significantly on contact duration, contact force, position and subject. When combining parameters wisely, like a contact pressure of 5 N, attaching the electrodes 5 minutes before performing a measurement and using a high frequency, the total impedances of the dry electrodes can even be lower than those of the Ag/AgCl dry gel electrodes.

In this work measurements have shown, that the usage of metal electrodes leads to lowest impedances and seem therefore to be the best choice for bioimpedance instrumentation. Additionally, the proposed metal electrodes can easily be produced and electrically contacted to measurement cables. However, it is important to note that this work did not analyse all electrical parameters, such as the occurring half-cell voltages. Unfortunately, these voltages are high when using metal electrodes (Grimnes and Martinsen 2008) and can therefore be challenging for bioimpedance measurement instrumentation. This parameter has comprehensively been analysed in other publications. For further decrease of the electrode-skin impedances, the surface areas of the selected electrodes might be increased.

The developed textile electrodes produced the highest impedance values. This is caused by the minor actual electrode surface, which is just represented by the weaved metal threads. In the magnified photograph of the electrode surface it can be seen, that these threads hardly touch the skin. Additionally, its characteristic of absorbing fluids prevents the arising sweat to work as electrolyte between the electrode and the skin. Nevertheless, there might be applications, in which the wearing comfort of textile electrodes is in focus.

## 5. Conclusion

For transferring bioimpedance measurement applications from research laboratories to real measurement environments, the electrode-skin impedances can be a major challenge. In contrast to experimental setups, in which gel electrodes might be acceptable, dry electrodes are often more useful and comfortable to the subject. Therefore, the development of five different kind of dry electrodes has been proposed and their impedance behaviors have been compared to commercially available Ag/AgCl dry gel electrodes.

The measurement results have shown, that metal electrodes and carbon rubber electrodes can be used to acquire the bioimpedance. However, to ensure a high reliability, the measurement setup has to meet certain parameters, like the electrode contact force or the measurement frequency.